\documentclass[
10pt,twocolumn,prl,aps,superscriptaddress,
floatfix,nobalancelastpage,preprintnumbers,reprint,
footinbib,noshowpacs,longbibliography
]{revtex4-2}
\usepackage[dvipsnames]{xcolor}
\usepackage[colorlinks=true,breaklinks=true]{hyperref}
\hypersetup{allcolors=[rgb]{0.0 0.0 0.70},linkcolor=blue}

\usepackage{orcidlink}
\usepackage{microtype}
\usepackage{float}
\usepackage{graphicx}
\usepackage{dcolumn}
\usepackage{bm}
\usepackage[normalem]{ulem}
\usepackage{esvect}
\usepackage{amsmath}
\usepackage{multirow}
\usepackage{amsfonts}
\usepackage{mathtools}
\usepackage{multirow}
\usepackage{bm}
\usepackage[utf8]{inputenc}
\usepackage{hyperref}
\usepackage{ulem}

\usepackage{xcolor}

\makeatletter

\newcommand{\checknextarg}{\@ifnextchar\bgroup{\gobblenextarg}{}}
\newcommand{\gobblenextarg}[1]{\,\mathrm{#1}\@ifnextchar\bgroup{\gobblenextarg}{}}
\makeatother

\begin{document}

\title{Unveiling Neutrino Nature with the Diffuse Supernova Background}

\author{Marco Manno\orcidlink{0009-0009-8622-8594}}
\email{marco.manno@unisalento.it}
\affiliation{%
Dipartimento di Matematica e Fisica ``Ennio De Giorgi,'' Universit\`a del Salento, 73100 Lecce, Italy
}
\affiliation{%
INFN-Istituto Nazionale di Fisica Nucleare, Sezione di Lecce, 73100 Lecce, Italy
}
\author{Pablo Mart{\'i}nez-Mirav{\'e}\orcidlink{0000-0001-8649-0546}}
\email{pablo.mirave@nbi.ku.dk}
\affiliation{%
Niels Bohr International Academy and DARK, Niels Bohr Institute, University of Copenhagen,
Blegdamsvej 17, 2100 Copenhagen, Denmark
}%
\author{Irene Tamborra\orcidlink{0000-0001-7449-104X}}
\email{tamborra@nbi.ku.dk}
\affiliation{%
Niels Bohr International Academy and DARK, Niels Bohr Institute, University of Copenhagen,
Blegdamsvej 17, 2100 Copenhagen, Denmark
}%

\date{\today}

\begin{abstract}
The true nature of neutrinos--whether Dirac or Majorana--is a foundational, unresolved question.  We demonstrate that the diffuse supernova neutrino background (DSNB) offers  an untapped avenue to resolve  this issue, provided  neutrino magnetic moments are $\gtrsim 10^{-14}\mu_B$. The intense magnetic fields characteristic of a subset of collapsing massive stars can trigger resonant chirality flips. This flavor conversion physics  alters DSNB fluxes in  measurable ways that depend  on the neutrino nature.
A $20$~yr combined exposure at  Hyper-Kamiokande loaded with gadolinium and JUNO can unravel this signature at $90\%$ ($99\%$) confidence if the fraction of magnetorotational events exceeds $12\%$ ($20\%$) of cosmic core collapses.
This result holds independent of the mass ordering and establishes  the DSNB as a critical gateway to  unveiling the true identity of the neutrino. 
\end{abstract}

\maketitle

{\em Introduction.}---Are neutrinos Dirac or Majorana particles?  This is a burning question in modern physics.
The nature of neutrinos  is closely related to the origin of neutrino mass and the conservation  of total lepton number~\cite{BahaBalantekin:2018ppj,Mohapatra:2006gs,deGouvea:2016qpx,Giunti:2007ry}.
Determining whether neutrinos are Dirac or Majorana particles is also key to understanding 
their role in generating the cosmic matter--antimatter asymmetry.

The primary experimental strategy to address this question is the search for neutrinoless double beta decay, $0\nu\beta\beta$~\cite{Furry:1939qr}. This process violates total lepton number by two units: its observation would imply a Majorana component of the neutrino mass~\cite{Schechter:1981hw,Rodejohann:2011mu}. No evidence for such a decay has yet been found~\cite{Gomez-Cadenas:2023vca}, but this  does not imply that neutrinos are Dirac particles.

The cosmic neutrino background (C$\nu$B) provides an alternative  probe into the nature of neutrinos,
with the  tritium capture rate being  twice as large for Majorana neutrinos  in standard cosmology. This is  because both relic helicity populations contribute, whereas only one does for Dirac neutrinos~\cite{Long:2014zva,Vitagliano:2019yzm}. 
PTOLEMY aims to detect the C$\nu$B~\cite{PTOLEMY:2019hkd}, which   may also be accessible to  upcoming neutrino telescopes~\cite{Herrera:2026pzj}, but such a measurement is extremely challenging and further complicated by the uncertain local clustering of relic neutrinos~\cite{deSalas:2017wtt,Lisanti:2014pqa,Hannestad:2009xu}.

Where else, then, could neutrino nature reveal itself? We argue that the answer may lie in a cosmic signal now approaching discovery: the diffuse supernova neutrino background (DSNB). The DSNB  is built up over cosmic history by neutrinos emitted during the  core collapse of massive stars~\cite{Bisnovatyi-Kogan:1982oyy,Krauss:1983zn,Domogatskii:1984a,Beacom:2010kk,Lunardini:2010ab,Mirizzi:2015eza,Ando:2023fcc,Tamborra:2024fcd}.  
A recent, model-independent binned analysis combining data from the pure-water and gadolinium-loaded periods of Super-Kamiokande disfavors the background-only hypothesis at $2.4 \sigma$~\cite{Sekiya:Neutrino2026SKSN}.
Furthermore, DSNB detection stands as a major target for the Jiangmen Underground Neutrino Observatory (JUNO), operating since August 2025~\cite{JUNO:2025gmd},  with Hyper-Kamiokande expected to join the search by the end of the decade~\cite{Hyper-Kamiokande:2025asb}. 

In this {\it Letter}, we establish the DSNB as a powerful diagnostic tool to distinguish between Dirac and Majorana neutrinos.  As neutrinos stream out of the core of collapsing massive stars,  flavor conversion molds their energy spectra~\cite{Lunardini:2012ne}. The vast majority of core-collapse supernovae  are expected to be neutrino-driven--see, e.g., Refs.~\cite{Janka:2025tvf,Raffelt:2025wty,Burrows:2020qrp,Mezzacappa:2005ju} for  reviews. However, a fraction of core-collapse supernovae  may be magnetorotational, i.e., stemming from  the collapse of rapidly
rotating stars harboring large magnetic fields~\cite{Muller:2020ard,Bisnovatyi-Kogan:2018vvk,Ardeljan:2004fq}. Neutrinos from  magnetorotational  collapses enhance the high-energy tail of the DSNB~\cite{Martinez-Mirave:2024zck}.
In addition, the strong magnetic fields of magnetorotational  collapses can  drive resonant neutrino chirality flips~\footnote{Note that this phenomenon is also referred to as resonant spin-flavor precession for  Majorana neutrinos  and resonant spin precession for Dirac neutrinos in the literature~\cite{Akhmedov:1988uk,Lim:1988tk,Akhmedov:2003fu,Ando:2002sk,Ando:2003pj,Ando:2003is,Abbar:2020ggq, Dvornikov:2011dv,Jana:2022tsa}. Following Ref.~\cite{Manno:2026ikk}, we use  ``magnetically-driven resonant conversion'' or ``B-res conversion'' instead.}, provided neutrinos possess magnetic dipole moments above $10^{-14}\,\mu_B$~\cite{Manno:2026ikk}.
Such magnetically-driven  resonant conversion   reshuffles active neutrino and antineutrino flavors for Majorana particles and populates opposite-chirality states that do not participate in Standard Model weak interactions   for Dirac particles (therefore evading detection). Regardless of the neutrino mass ordering, we show that magnetically-driven flavor conversion imprints distinct signatures on the DSNB depending on neutrino nature, with such differences becoming increasingly pronounced for neutrino energies $\gtrsim 15$~MeV.

{\em Diffuse supernova neutrino background}.--- Following the framework of Ref.~\cite{Martinez-Mirave:2024zck}, we consider  the contributions to the DSNB from neutrino-driven core-collapse supernovae ($\nu$-SN)  and magnetorotational (MR) collapses:
\begin{align}
 \Phi_{\rm DSNB}
 =
 (1-f_{\rm MR})\Phi_{\nu-{\rm SN}}
 +f_{\rm MR}\Phi_{\rm MR}\, ;
 \label{eq:total-dsnb-flux}
\end{align} 
$\Phi_{\nu-{\rm SN}}$ ($\Phi_{\rm MR}$)  is the diffuse flux from successful neutrino-driven  supernovae   and black hole forming collapses (magnetorotational collapses);  the  fraction of magnetorotational collapses, $f_{\rm MR}$, is  assumed  constant over redshift.

The DSNB contribution from neutrino-driven supernovae is given by
\begin{align}
 \label{eq:ccsn-diffuse-flux}
 \Phi_{\nu-{\rm SN}}(E)
 =
 &c
 \int_{8 M_\odot}^{125 M_\odot} dM
\\
 &\times
 \int_{0}^{z_{\rm max}} dz\,
 \frac{\mathcal R(z,M)}{H(z)}
 \varphi_\nu\!\left[M,E_\nu(1+z)\right]\, , \nonumber
\end{align}
where $E_\nu$ is the observed neutrino energy and  $H(z)$ denotes the Hubble expansion rate. The time-integrated neutrino spectrum, $\varphi_\nu[M,E_\nu(1+z)]$, accounts for flavor conversion and is computed for each supernova with zero-age main-sequence mass $M$  at redshift $z$.  The core-collapse supernova rate, $\mathcal R(z,M)$, 
combines the Salpeter initial mass function~\cite{Salpeter:1955it} with the fiducial star formation history~\cite{Horiuchi:2008jz}.  We assume $z_{\rm max}=5$ since any contribution from higher redshifts would be negligible~\cite{Ando:2004hc} and adopt the cosmological parameters of Refs.~\cite{Abdalla:2022yfr,Planck:2018vyg}.  
Within the neutrino-driven population, we assume that some supernovae explode successfully and a fraction   fails to explode ($0\leq f_{\rm BH}\leq0.47$), forming a black hole~\cite{Kresse:2020nto,Moller:2018kpn,Ertl:2015rga,Sukhbold:2015wba}.

The neutrino emission from  successful neutrino-driven explosions is computed using  spherically symmetric  hydrodynamical simulations without muons~\cite{Mirizzi:2015eza,Garching}, with the Lattimer
and Swesty nuclear equation of state, with an incompressibility modulus of $K = 220$~MeV~\cite{Lattimer:1991nc}.  We adopt  the $11.2 M_\odot$ ($27 M_\odot$) model as representative of the neutrino-driven core collapses with mass $\lesssim 15 M_\odot$ ($> 15 M_\odot$).
As representative of black hole forming collapses, we use the $40M_\odot$ model (model s4027b2)~\cite{Mirizzi:2015eza,Garching}.  

We assume that the three-dimensional   model with mass of $13 M_\odot$~\cite{Manno:2026ikk,obergaulinger-aloy,Obergaulinger:2021omt} is   representative of the magnetorotational population.
We compute the neutrino fluxes averaging  over the emission directions and   accounting for flavor conversion for Dirac and Majorana (anti)neutrinos; cf.~Eqs.~\eqref{eq:fluxes-majorana} and \eqref{eq:fluxes-dirac}.  
$\Phi_{\rm MR}$ is computed as in Eq.~\eqref{eq:ccsn-diffuse-flux}, with the mass integral taken over $5M_\odot\leq M\leq125M_\odot$ to account for chemically homogeneous, rapidly rotating progenitors~\cite{Martinez-Mirave:2024zck}. We assume $0\leq f_{\rm MR}\leq0.26$~\cite{Frohmaier:2020pec,Martinez-Mirave:2024zck}.

{\em Flavor conversion.}---Neutrinos change their  flavor as they propagate in the core of  collapsing massive stars. At high densities,  flavor conversion is dominated by neutrino-neutrino interaction--see Refs.~\cite{Tamborra:2020cul,Tamborra:2024fcd,Volpe:2023met,Johns:2025mlm} for recent reviews. Because of the uncertainties plaguing the modeling of neutrino collective effects, we neglect them and  focus  on matter- and magnetically-induced flavor conversion. 

At distances larger than $\mathcal{O}(10^3)$~km, coherent forward scattering on matter dominates the flavor evolution, leading to the Mikheyev--Smirnov--Wolfenstein (MSW) resonances~\cite{Wolfenstein:1977ue,Mikheyev:1985zog,Dighe:1999bi}. The outcome of MSW conversion depends on the neutrino mass ordering, normal (NO) or inverted (IO), but is identical for Dirac and Majorana neutrinos; its impact on the DSNB has been studied extensively~\cite{Lunardini:2012ne}. 

If neutrinos have non-zero magnetic moment~\cite{Giunti:2014ixa}, they can additionally undergo resonant transitions involving a chirality flip in  magnetorotational collapses~\cite{Cisneros:1971,Fujikawa:1980yx,Voloshin:1986ty,Okun:1986na,
Schechter:1981hw,Manno:2026ikk}.
For Majorana particles,  neutrinos can convert into antineutrinos and vice versa.  For Dirac particles, the chirality flipping  interactions transform the initial neutrinos or antineutrinos into opposite-chirality states  that do not participate in Standard Model weak interactions and are not  detectable. 
We rely on Ref.~\cite{Manno:2026ikk} for an overview of the magnetically-driven resonant conversion conditions  for Majorana neutrinos, whereas we present the  conditions for magnetically-driven resonant conversions for Dirac neutrinos in the Supplemental Material. 

\begin{figure}[t]
    \centering
    \includegraphics[width=\columnwidth]{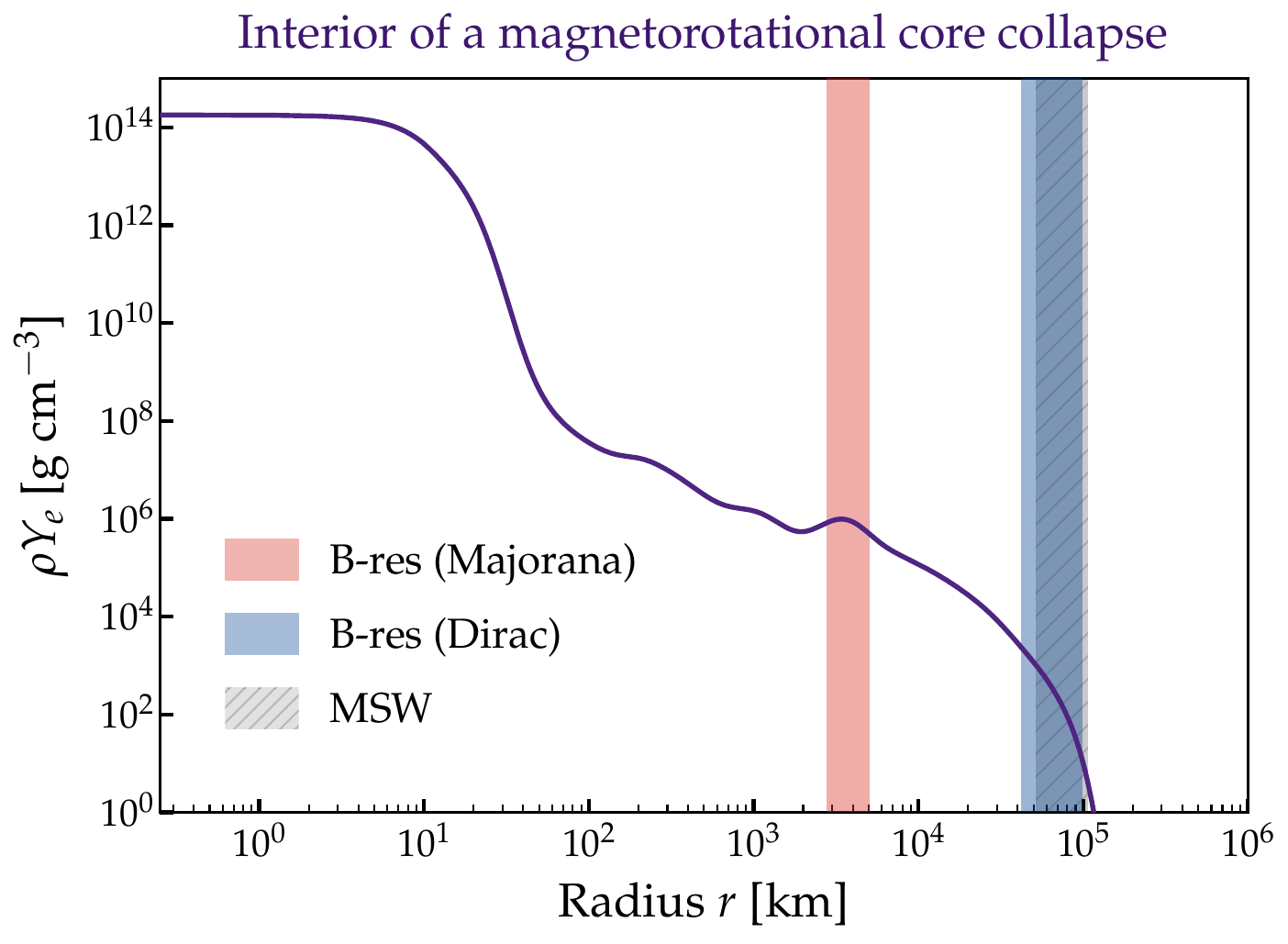}
    \caption{Radial ranges of the MSW and magnetically-driven resonances for our benchmark magnetorotational model, assuming NO. The solid line shows the profile of $\rho Y_e$ along the polar direction at $1$~s after bounce, where $\rho$ is the baryon mass density and $Y_e$ is the electron fraction; the corresponding ones in IO are approximately comparable. The resonance locations are evaluated for a representative neutrino energy, $E_\nu=14.4$~MeV, corresponding to the time-averaged mean energy across all neutrino species~\cite{Manno:2026ikk}. 
    The red and blue vertical bands indicate the regions for magnetically-driven resonant conversion  (B-res)  for Majorana and Dirac neutrinos, respectively. The hatched band marks the  MSW region. The magnetically-driven resonant conversion  occurs at larger radii for Dirac neutrinos than for Majorana neutrinos.}
    \label{fig:resonance-regions}
\end{figure}

Figure~\ref{fig:resonance-regions} shows the radial regions affected by MSW and magnetically-driven resonant conversion in our magnetorotational model.
The magnetically-driven resonant conversion for Majorana particles occurs  at a few $10^3$~km, while   resonant flavor conversion    at $10^4$--$10^5$~km  for Dirac neutrinos. Therefore, the MSW and magnetically-driven  conversion regions  may overlap for Dirac neutrinos. 

We find that all magnetically-driven resonances  are adiabatic for  Dirac and Majorana particles if the neutrino magnetic moments are $\gtrsim 10^{-14}\,\mu_B$; the adiabaticity threshold may vary with neutrino energy and post-bounce time, as discussed in the Supplemental Material.
The range of magnetic moments ensuring adiabatic conversion lies below the most stringent  bounds: $\mu\lesssim 10^{-12}\,\mu_B$~\cite{Giunti:2014ixa,ParticleDataGroup:2024cfk}.
Differences between Dirac and Majorana particles can also arise for smaller magnetic moments, when only some crossings are adiabatic. However, we focus on  the  adiabatic scenario; cf.~Eqs.~\eqref{eq:dirac-no-antinu-general} and \eqref{eq:dirac-io-antinu-general}  of the Supplemental Material (with jump probabilities set to zero) for Dirac neutrinos, and 
Eqs.~(21) and (23) of Ref.~\cite{Manno:2026ikk} for Majorana neutrinos, and  the  MSW-only scenario. 

After accounting for flavor conversion,  the $\bar\nu_e$ flux  at Earth for each collapsing massive star is
\begin{equation}
 \varphi_{\bar\nu_e}^{\rm M}=
 \begin{cases}
  ( |U_{e1}|^2+|U_{e3}|^2 )\varphi_{\bar\nu_x}^0
  +|U_{e2}|^2\varphi_{\nu_e}^0
  & \mathrm{NO}\\
  |U_{e1}|^2\varphi_{\bar\nu_e}^0
  +( |U_{e2}|^2+|U_{e3}|^2 )\varphi_{\bar\nu_x}^0
  & \mathrm{IO}
 \end{cases}\, ,
 \label{eq:fluxes-majorana}
\end{equation}
and
\begin{equation}
 \varphi_{\bar\nu_e}^{\rm D}=
 \begin{cases}
  |U_{e1}|^2\varphi_{\bar\nu_e}^0+|U_{e3}|^2\varphi_{\bar\nu_x}^0
  & \mathrm{NO}\\
  |U_{e2}|^2\varphi_{\bar\nu_x}^0
  & \mathrm{IO}
 \end{cases}
 \label{eq:fluxes-dirac}\, ,
\end{equation}
where M and D stand for Majorana and Dirac particles.
The lepton mixing matrix elements are $|U_{e1}|^2 = \cos^2\theta_{12}\cos^2\theta_{13}$, $|U_{e2}|^2 = \sin^2\theta_{12}\cos^2\theta_{13}$, and $|U_{e3}|^2 = \sin^2\theta_{13}$. We adopt the best-fit mixing angles and mass splittings of Ref.~\cite{deSalas:2020pgw}. The superscript $0$ denotes the  spectrum  before flavor conversion, and $\nu_x=\bar\nu_x= \bar\nu_{\mu, \tau}$.

\begin{figure}[t]
    \centering
    \includegraphics[width=\columnwidth]{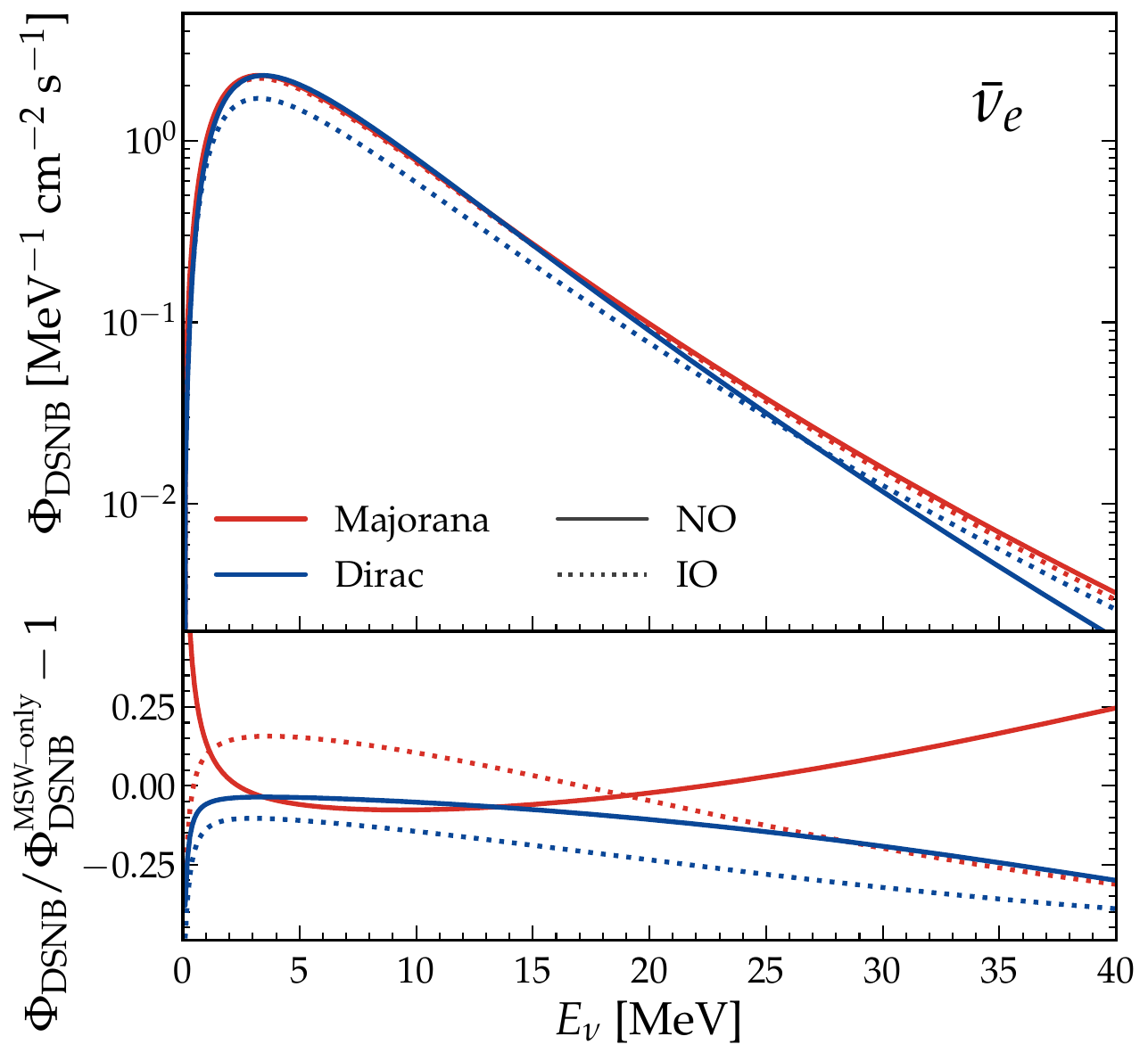}
    \caption{
    {\it Top panel:} Electron antineutrino DSNB flux  for Majorana (red) and Dirac (blue) neutrinos as a function of the neutrino energy, obtained  for  $f_{\rm MR}=0.1$ and $f_{{\rm BH}}=0.3$ for representative purposes. The solid (dotted) lines assume NO (IO). {\it Bottom panel:}   Deviations of the $\bar\nu_e$ DSNB flux in the scenario accounting for magnetically-driven resonant conversion relative to the MSW-only prediction. 
    Magnetically-driven conversion produces opposite  deviations of the spectra above $15$~MeV for Majorana and Dirac neutrinos in NO, with respect to the MSW-only scenario. The high-energy tail of the spectra is depleted in IO by different amounts for Dirac and Majorana antineutrinos.
    }
    \label{fig:dsnb-flux-no}
\end{figure}

Figure~\ref{fig:dsnb-flux-no} (upper panel) shows the $\bar\nu_e$ DSNB flux defined as in Eq.~\eqref{eq:total-dsnb-flux}. For  magnetorotational core collapses, we account for magnetically-driven resonant  conversions   for Majorana and Dirac antineutrinos (Eqs.~\ref{eq:fluxes-majorana} and \ref{eq:fluxes-dirac}), while for neutrino-driven core collapses, we only consider MSW transitions (see Eq.~\ref{eq:msw-only-antinu-fluxes}, since magnetic fields are negligible).
We use the representative choice $f_{\rm MR}=0.1$ and $f_{{\rm BH}}=0.3$. 
We find an excess (deficit) relative to the standard MSW-only scenario for Majorana (Dirac) particles in NO, as visible from the bottom panel of Fig.~\ref{fig:dsnb-flux-no}; in IO, the high-energy tail of the spectra tends to be depleted with respect to the MSW-only scenario, with visible differences for Dirac and Majorana particles.  

\begin{figure}[t]
    \centering
    \includegraphics[width=\columnwidth]{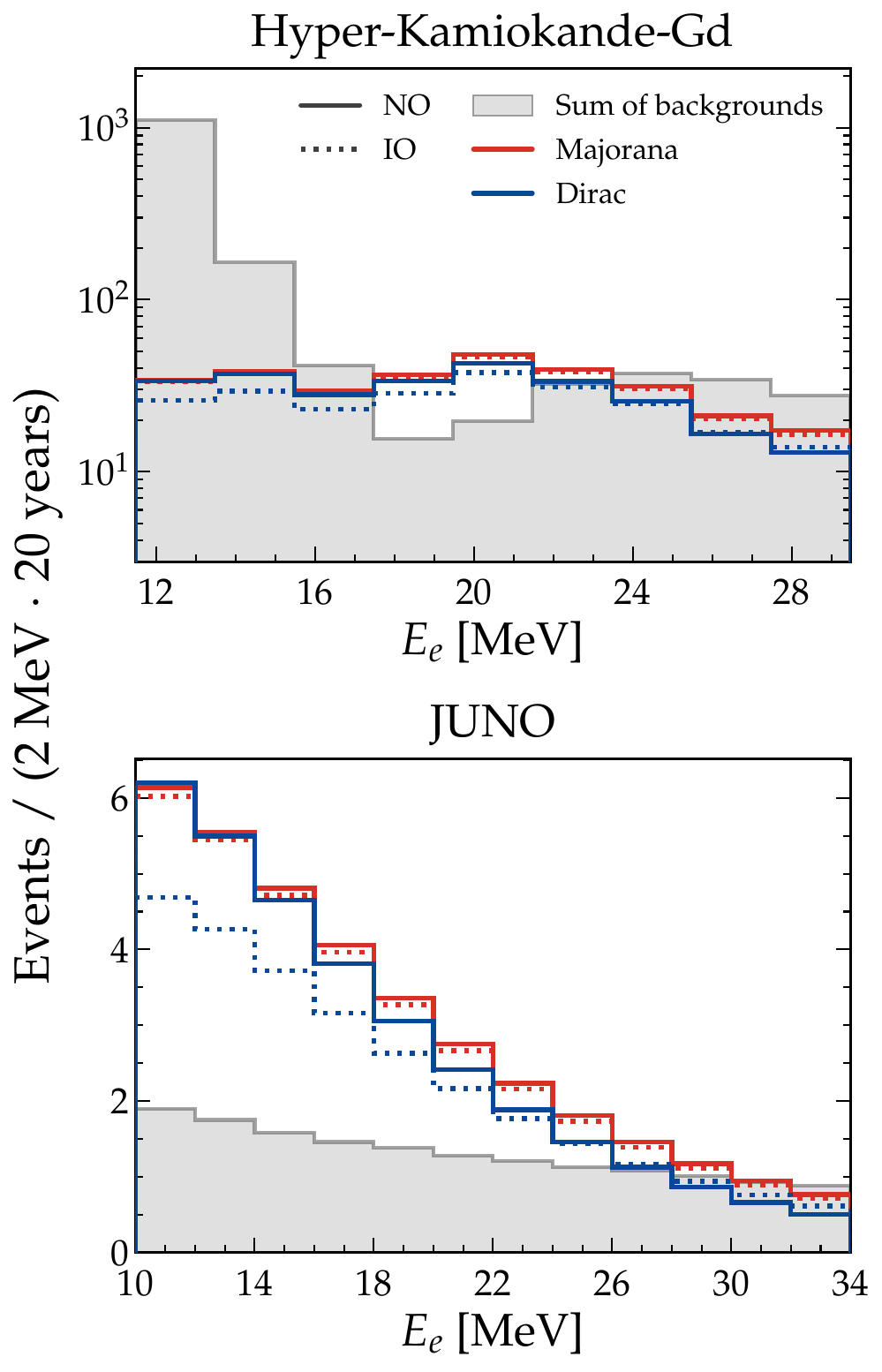}
    \caption{Expected event rates for Hyper-Kamiokande-Gd (top panel) and JUNO (bottom panel) after $20$ yr of data taking, obtained for $f_{\rm MR}=0.1$ and $f_{{\rm BH}}=0.3$ for representative purposes. The gray histograms show the total detector backgrounds, while the red (blue) line shows the DSNB signal for Majorana (Dirac) particles. Solid (dotted) lines denote NO (IO).}
    \label{fig:event-rates}
\end{figure}

{\em Expected event rate.}---Hyper-Kamiokande loaded with gadolinium (Gd) and JUNO will detect  DSNB $\bar\nu_e$'s  primarily through inverse beta decay ($\bar\nu_e+p\to e^++n$).  
The expected number of signal events in the reconstructed positron energy bin $i$ is
\begin{align}
 N_i^{\rm sig}
 =
 &\tau N_t
 \int_{E_\nu^{\rm th}}^\infty dE_\nu\,
 \Phi_{\bar\nu_e}(E_\nu)\sigma_{\rm IBD}(E_\nu,E'_e)
 \nonumber\\
 &\times
 \int_{E_i}^{E_{i+1}} dE_e\,
 \varepsilon(E_e)K(E'_e,E_e)\, ,
 \label{eq:detected-spectrum}
\end{align}
where $\tau$ is the exposure time, $N_t$ is the number of free proton targets, $E_\nu^{\rm th}=1.806$~MeV is the energy threshold, and $E'_e\simeq E_\nu-m_n+m_p$ with $m_n$ and $m_p$ being the neutron and proton masses. Here, $E_\nu$, $E'_e$, and $E_e$ are the true antineutrino, true positron, and reconstructed positron energies,
respectively.  The lower and upper limits of the reconstructed energy bin $i$ are  $E_i$ and $E_{i+1}$. The inverse-beta decay cross section, $\sigma_{\rm IBD}$, is parameterized as in 
Refs.~\cite{Strumia:2003zx,Ricciardi:2022pru}. The detector response function $K$ is assumed to be the same for both detectors and is modeled following  Appendix A  of  Ref.~\cite{MacDonald:2024vtw}.

For Hyper-Kamiokande-Gd, we double the energy-dependent efficiency, $\varepsilon(E_e)$, with respect to the  Hyper-Kamiokande one~\cite{Hyper-Kamiokande:2018ofw},  motivated by the improved neutron-tagging performance~\cite{Santos:2025phd}.
We model the backgrounds  following Ref.~\cite{Santos:Neutrino2026HKSN} and do not consider the additional background reduction expected from gadolinium loading. We assume $N_t=2.5\times10^{34}$ free proton targets and reconstructed positron energies between $11.48$ and $29.48$~MeV. 

For JUNO~\cite{JUNO:2015zny,JUNO:2021vlw}, we assume constant efficiency, $\varepsilon=0.8$, projected to be achievable with pulse-shape discrimination~\cite{Cheng:2023zds}, and account for  backgrounds following Ref.~\cite{JUNO:2022lpc}. We use $N_t=1.2\times10^{33}$ free proton targets and reconstructed positron energies between $10$ and $34$~MeV.

Figure~\ref{fig:event-rates} shows the corresponding event spectra for Hyper-Kamiokande-Gd   and JUNO after $20$~yr of data taking. The gray histograms show the total detector background for each energy bin, and the solid lines represent the DSNB event spectra for Dirac and Majorana neutrinos.  We can clearly see that the DSNB event spectra for Dirac and Majorana particles are distinguishable from each other,  in both Hyper-Kamiokande-Gd and JUNO,  independent of the mass ordering. A comparison of these results with the ones expected in the MSW conversion scenario is provided in  the Supplemental Material.

We expect that a joint statistical analysis of the  event rates of  Hyper-Kamiokande-Gd and JUNO  should increase the statistical significance, reducing the  impact of detector backgrounds and astrophysical degeneracies. Although DUNE can further reduce astrophysical degeneracies~\cite{Moller:2018kpn}, we find that $\nu_e$'s are less sensitive to the Dirac vs.~Majorana discrimination.
Constraints from coherent elastic neutrino--nucleus scattering detectors  may also provide complementary
constraints~\cite{Suliga:2021hek}.

{\em Statistical analysis}.---We employ a frequentist approach and define the test statistic as follows:
\begin{align}
\Delta\chi^2={}&\min_{\xi;\eta_{d,k}}
\Bigg\{2\sum_{d,i}\left[
 N_{d,i}^{\rm test}-N_{d,i}^{\rm true}
 +N_{d,i}^{\rm true}\ln\!\left(
 \frac{N_{d,i}^{\rm true}}{N_{d,i}^{\rm test}}
 \right)\right]
 \nonumber\\
 &+\left(\frac{\xi}{\sigma_{\mathcal{R}}}\right)^2
 +\sum_{d,k}\left(\frac{\eta_{d,k}}{\sigma_{d,k}}\right)^2\Bigg\}\, ,
 \label{eq:statistical-analysis}
\end{align}
where $d$ stands for Hyper-Kamiokande-Gd  or JUNO, $i$ runs over the energy bins of the corresponding neutrino telescope, and $k$ labels its background components.  The event rates under the test and true hypotheses are $N_{d,i}^{\rm test}=(1+\xi)S_{d,i}^{\rm test} +\sum_k(1+\eta_{d,k})B_{d,k,i}$ and $N_{d,i}^{\rm true}=S_{d,i}^{\rm true}+\sum_k B_{d,k,i}$; here $S_{d,i}$ denotes the  signal event rate and $B_{d,k,i}$ is the event rate for the background component $k$ in the detector $d$ and energy bin $i$.

Each background component $k$ at the detector $d$  has an independent nuisance parameter $\eta_{d,k}$. We consider a $5\%$ uncertainty for the background of atmospheric neutrinos that do not undergo neutral-current quasi-elastic interactions (non-NCQE) in Hyper-Kamiokande~\cite{Santos:2025phd} and a $20\%$ uncertainty for all other background components in both detectors. 
The nuisance parameter $\xi$ accounts for the uncertainty in the supernova rate normalization. We assume $\sigma_{\mathcal{R}}=5\%$~\cite{Lien:2010yb} and marginalize over  $f_{{\rm BH}} \in [0, 0.47]$.

We evaluate $\Delta\chi^2$ for $f_{\rm MR}$ in the range $[0,0.26]$.
First we assume neutrinos are Dirac particles and test the hypothesis against the predictions for Majorana particles; then, we swap the assumptions. 

\begin{figure}[t]
    \centering
    \includegraphics[width=\columnwidth]{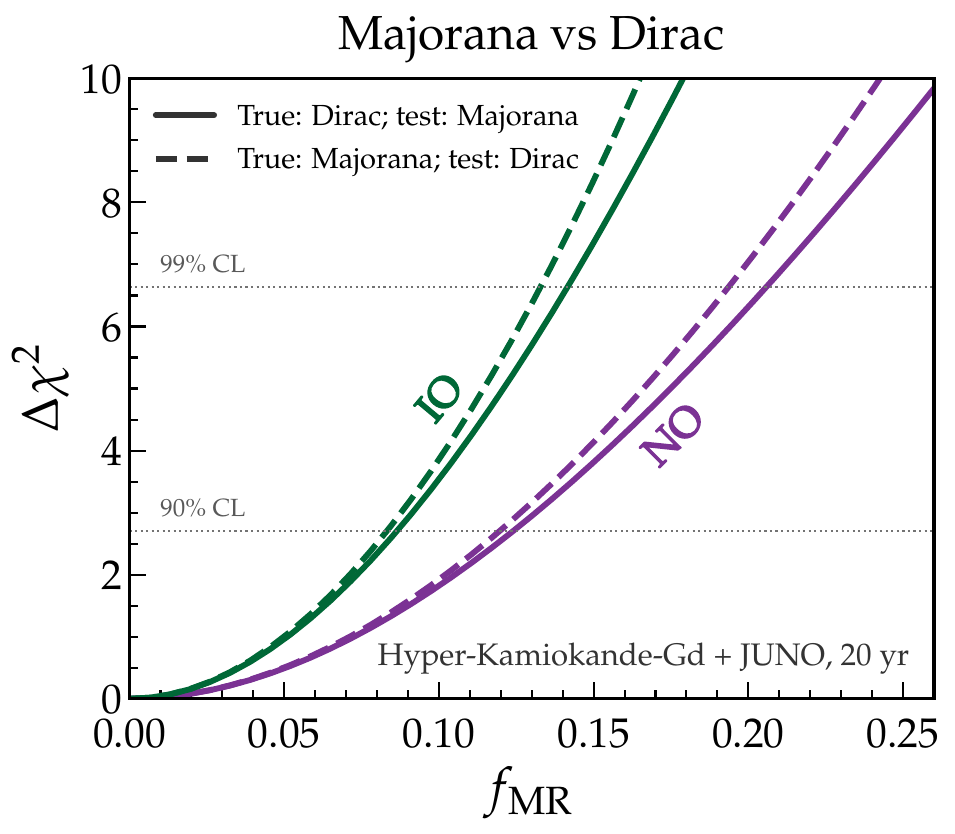}
    \caption{Expected sensitivity to  Dirac vs.~Majorana neutrinos using $20$~yr of DSNB data from Hyper-Kamiokande-Gd  and JUNO. The purple and green lines denote NO and IO, respectively. Solid lines correspond to assuming Dirac neutrinos as the true hypothesis and testing Majorana neutrinos, while the dashed lines show the reverse test. The horizontal lines mark the one-dimensional $90\%$ and $99\%$
    confidence levels. The nature of neutrino masses can be discriminated with $90\%$ ($99\%$) confidence  for $f_{\rm MR}\gtrsim 0.12$ ($0.19$--$0.21$) in NO and $f_{\rm MR}\gtrsim 0.09$ ($0.13$--$0.14$) in IO. }
    \label{fig:sensitivity-majorana-dirac}
\end{figure}

Figure~\ref{fig:sensitivity-majorana-dirac} shows the projected sensitivity, after $20$~yr of combined data from  Hyper-Kamiokande-Gd and JUNO, to discriminate the Dirac or Majorana nature of neutrinos as a function of $f_{\rm MR}$ for both mass orderings.
We can see that we will be able to discriminate between Dirac and Majorana particles at $90\%$ confidence   for $f_{\rm MR}\gtrsim 0.12$ in NO and $f_{\rm MR}\gtrsim 0.08$--$0.09$ in IO;  moreover, we could infer the nature of neutrinos with $99\%$ confidence, if  $f_{\rm MR} \gtrsim 0.19$--$0.21$ ($f_{\rm MR}\gtrsim 0.13$--$0.14$) in NO (IO). 
These results show that magnetically-driven resonant flavor conversion could be a key ingredient in neutrino physics that  could  allow us to distinguish Dirac from Majorana particles within the currently allowed range for $f_{\rm MR}$. 

The supernova rate  uncertainties  and  the detector backgrounds have the largest impact on our analysis. The atmospheric non-NCQE background dominates  above $18$~MeV  where the Dirac and Majorana predictions differ the most~\cite{Santos:Neutrino2026HKSN}. 
Hence, it is essential to reduce such uncertainties, and employ gadolinium loading in  Hyper-Kamiokande.  Although we marginalize over $f_{\rm BH}$, this parameter affects the high-energy tail of the DSNB~\cite{Lunardini:2009ya} and may  weaken the Dirac vs.~Majorana discrimination if $f_{\rm BH}$ were to be close to the upper limit used in this work. We  retain the smallest $\Delta\chi^2$ over $0\leq f_{\rm BH} \leq 0.47$, making the quoted sensitivity conservative.

For neutrino magnetic moments below $10^{-14} \mu_B$, magnetically-driven resonant conversion would strongly depend on neutrino energy and progenitor properties; therefore we conservatively assume that only MSW transitions would take place.
This scenario would be  distinguishable from the case where magnetically-induced resonant flavor conversion takes place, strengthening our confidence in determining the nature of neutrino masses. 

{\em Conclusions.}---The DSNB is  transitioning from a  theoretical prediction to an imminent  observable. Beyond probing the core-collapse supernova population and offering insight into New Physics  (though see Ref.~\cite{MacDonald:2024vtw}),  the DSNB holds the potential to address a  fundamental question: the  nature of the neutrino. Driven by magnetically-driven  resonant flavor conversion, distinct DSNB event rates emerge  for Majorana and Dirac neutrinos above $15$~MeV, if their magnetic dipole moments are $\gtrsim 10^{-14} \mu_B$.

We demonstrate that a $20$~yr combined exposure of  Hyper-Kamiokande-Gd and JUNO can distinguish between Dirac and Majorana neutrinos  at  $90\%$  ($99\%$) confidence,  provided that the fraction of magnetorotational collapses exceeds $12\%$ ($20\%$)--a result that holds independent of the neutrino mass ordering.
Achieving such  sensitivity relies  on tightening constraints on the cosmic   core-collapse supernova rate (that we assume to be measured up to $5\%$ uncertainty), further mitigating the expected experimental backgrounds, and successfully implementing gadolinium loading in  Hyper-Kamiokande. These advances are already central objectives of future plans to measure the  DSNB. Our findings highlight the DSNB as a powerful, high-energy frontier for pinning down the fundamental nature of the neutrino.

{\em Acknowledgments.}---We thank Miguel
Angel {\'A}loy and Martin Obergaulinger for  kindly sharing
the data of the magnetorotational collapse model employed in this work. M.M.~acknowledges financial support from the research
project TAsP (Theoretical Astroparticle Physics), funded by the Istituto
Nazionale di Fisica Nucleare (INFN).
In Copenhagen, this project has received support from the European Union 
(ERC, ANET, Project No.~101087058) and the Elite Research Prize from the Danish Ministry of Higher Education and Science (Project No.~3142-00074B). Views and opinions expressed are those of the authors only and do not necessarily reflect those of the European Union or the European Research Council. Neither the European Union nor the granting authority can be held responsible for them.

\bibliography{bibliography}

\clearpage
\onecolumngrid
\appendix

\newpage
\clearpage

\setcounter{equation}{0}
\setcounter{figure}{0}
\setcounter{table}{0}
\setcounter{page}{1}
\makeatletter
\renewcommand{\theequation}{S\arabic{equation}}
\renewcommand{\thefigure}{S\arabic{figure}}
\renewcommand{\thetable}{S\arabic{table}}
\renewcommand{\thepage}{S\arabic{page}}

\begin{center}
\textbf{\large Supplemental Material\\[0.8ex]
Unveiling Neutrino Nature with the Diffuse Supernova Background }
\end{center}

In this Supplemental Material, we explore the physics of magnetically-driven resonant conversion for Dirac neutrinos. We also contrast the flavor configuration expected in the presence of magnetically-driven resonant conversion with  the scenario where only MSW conversion occurs for both Dirac and Majorana neutrinos.   

\bigskip

\section{A.~Magnetically-driven resonant conversion of Dirac neutrinos}

Magnetically-driven resonant conversion of  Majorana neutrinos has been explored extensively in Refs.~\cite{Akhmedov:1988uk,Lim:1988tk,Akhmedov:2003fu,Ando:2002sk,Ando:2003pj,Ando:2003is,Abbar:2020ggq, Dvornikov:2011dv,Jana:2022tsa,Manno:2026ikk}. 
Here, we derive the analogous resonance conditions for Dirac neutrinos and  apply such resonance conditions to  the magnetorotational model adopted in the main text. We  determine the resonance locations, estimate the adiabaticity of flavor conversion, and compute the oscillated  fluxes expected at Earth.

\subsection{A.1~Resonance conditions}
We describe the flavor and chirality evolution for Dirac neutrinos and antineutrinos through the $6\times6$ density matrices, $\varrho_{\rm D}^{\nu}$ and $\varrho_{\rm D}^{\bar\nu}$. In the flavor basis, each matrix is block diagonal with two $3\times3$ blocks. For neutrinos, the first block describes the active states $\nu_{\alpha L}$ and the  second corresponds to opposite-chirality states $\nu_{\alpha R}$. For antineutrinos, the  ordering is $\bar\nu_{\alpha R}$ followed by $\bar\nu_{\alpha L}$, with $\alpha = e, \mu, \tau$. Here, opposite-chirality is defined with respect to the corresponding active state that participates in Standard Model weak interactions.

The radial evolution of the density matrices is given by
\begin{equation}
 i\frac{d\varrho_{\rm D}^{a}}{dr}
 =[\mathcal H_{\rm D}^{a},\varrho_{\rm D}^{a}],
 \qquad a=\nu,\bar\nu\, .
 \label{eq:dirac-evolution}
\end{equation}
With this block ordering, the Hamiltonians are
\begin{equation}
 \mathcal H_{\rm D}^{\nu}=
 \begin{pmatrix}
  H_{\rm vac}+H_{\rm mat} & B_\perp\boldsymbol{\mu}_{\rm D}\\
  B_\perp\boldsymbol{\mu}_{\rm D}^{\dagger} & H_{\rm vac}
 \end{pmatrix}\, ,
 \qquad
 \mathcal H_{\rm D}^{\bar\nu}=
 \begin{pmatrix}
  H_{\rm vac}^{*}-H_{\rm mat} & B_\perp\boldsymbol{\mu}_{\rm D}^{*}\\
  B_\perp\boldsymbol{\mu}_{\rm D}^{T} & H_{\rm vac}^{*}
 \end{pmatrix}\, .
 \label{eq:dirac-hamiltonians}
\end{equation}
Here $B_\perp$ denotes the magnetic field component perpendicular to the radial direction of propagation of neutrinos emitted from the  3D magnetorotational model.
The $3\times3$ matrix $\boldsymbol{\mu}_{\rm D}$ is the magnetic-moment matrix of Dirac neutrinos in flavor space. Its diagonal elements couple states of the same flavor and opposite chirality and are intrinsic magnetic moments (absent for Majorana neutrinos), whereas its off-diagonal elements are transition magnetic moments coupling different flavors.

We define $\omega_{\rm H}=\Delta m^2_{31}/(2E_\nu)$ and $\omega_{\rm L}=\Delta m^2_{21}/(2E_\nu)$, where $\Delta m^2_{ij}=m_i^2-m_j^2$. Thus, $\omega_{\rm H}>0$ ($<0$) in NO (IO), whereas $\omega_{\rm L}>0$ in both cases. The vacuum Hamiltonian is 
\begin{equation}
 H_{\rm vac}=U\operatorname{diag}(0,\omega_{\rm L},\omega_{\rm H})U^\dagger\, ,
\label{eq:dirac-vacuum-hamiltonian}
\end{equation}
where $E_\nu$ is the neutrino energy and $U$ is the lepton mixing matrix, parametrized as in Ref.~\cite{ParticleDataGroup:2024cfk} in terms of the three mixing angles $\theta_{12}$, $\theta_{13}$, and $\theta_{23}$ and a complex phase $\delta$. For the magnetorotational model considered below, $\delta$ is expected to have a negligible impact on flavor conversion because the $\mu$- and $\tau$-flavor emission spectra are identical~\cite{Balantekin:2007es}. We adopt the best-fit values of the mixing angles and mass splittings from Ref.~\cite{deSalas:2020pgw}.

The Hamiltonian term accounting for interactions with matter is
\begin{equation}
 H_{\rm mat}=\operatorname{diag}(V_e+V_n,V_n,V_{\mu\tau}+V_n)\, .
 \label{eq:dirac-matter-hamiltonian}
\end{equation}
Here $V_e=C\rho Y_e$ and $V_n=-C\rho(1-Y_e)/2$ arise from coherent forward scattering on electrons and neutrons, respectively, with $C=\sqrt{2}G_{\rm F}/m_N$. The quantities $G_{\rm F}$, $m_N$, $\rho$, and $Y_e$ denote the Fermi constant, nucleon mass, baryon mass density, and electron fraction, respectively. The term $V_{\mu\tau}$ accounts for the difference between the effective matter potentials experienced by $\nu_\mu$ and $\nu_\tau$ due to radiative corrections~\cite{Botella:1986wy}. No matter term appears in the opposite-chirality blocks of Eq.~\eqref{eq:dirac-hamiltonians}, because these states do not participate in Standard Model weak interactions.

We work in the rotated basis introduced in Ref.~\cite{Akhmedov:2003fu}, obtained by rotating the $\mu$--$\tau$ sector through the mixing angle $\theta_{23}$. The rotated states are $\nu'_{\mu}=\cos\theta_{23}\nu_\mu-\sin\theta_{23}\nu_\tau$ and $\nu'_{\tau}=\sin\theta_{23}\nu_\mu+\cos\theta_{23}\nu_\tau$. We apply the same rotation to both chiralities and to neutrinos and antineutrinos.

The resonance locations are identified by equating the relevant diagonal entries of the Hamiltonians in Eq.~\eqref{eq:dirac-hamiltonians}. For $a,b=e,\mu',\tau'$, the active-active and active-opposite-chirality level crossing conditions are
\begin{equation}
 \begin{aligned}
 (H_{\rm vac}+H_{\rm mat})_{aa}
 &=(H_{\rm vac}+H_{\rm mat})_{bb}
 &\nu_{aL}&\leftrightarrow\nu_{bL}\, ,\\
 (H_{\rm vac}^{*}-H_{\rm mat})_{aa}
 &=(H_{\rm vac}^{*}-H_{\rm mat})_{bb}
 &\bar\nu_{aR}&\leftrightarrow\bar\nu_{bR}\, ,\\
 (H_{\rm vac}+H_{\rm mat})_{aa}
 &=(H_{\rm vac})_{bb},
 &\nu_{aL}&\leftrightarrow\nu_{bR}\, ,\\
 (H_{\rm vac}^{*}-H_{\rm mat})_{aa}
 &=(H_{\rm vac}^{*})_{bb},
 &\bar\nu_{aR}&\leftrightarrow\bar\nu_{bL}\, .
 \end{aligned}
 \label{eq:dirac-level-crossing-conditions}
\end{equation}
The first two lines of Eq.~\eqref{eq:dirac-level-crossing-conditions} describe active--active crossings and yield the approximate conditions for the standard MSW resonances~\cite{Dighe:1999bi}:
\begin{equation}
 \begin{aligned}
 \mathrm{MSW(H)}:\quad
 C\rho Y_e&\simeq |\omega_{\rm H}|\cos2\theta_{13}&
 \nu_{eL}&\leftrightarrow\nu'_{\tau L}\quad({\rm NO})\ \;\;\; \mathrm{and}\;\;\;
 \bar\nu_{eR}&\leftrightarrow\bar\nu'_{\tau R}\quad({\rm IO})\, ,\\
 \mathrm{MSW(L)}:\quad
 C\rho Y_e&\simeq \omega_{\rm L}\cos2\theta_{12}&
 \nu_{eL}&\leftrightarrow\nu'_{\mu L}\quad({\rm NO,IO})\, .
 \end{aligned}
 \label{eq:dirac-msw-resonances}
\end{equation}
The last two lines of Eq.~\eqref{eq:dirac-level-crossing-conditions} describe crossings between an active state and an opposite-chirality state. The resulting resonant conversions are magnetically-driven and labeled B-res (L,H)$_\alpha$. Here, the subscript identifies the flavor involved in the crossing (primes are omitted), while $x$ denotes the non-electron sector $\mu'-\tau'$.

For $e\leftrightarrow\mu'$ crossings,
 the condition for magnetically-driven resonant conversions of $\nu_{eL}\leftrightarrow\nu'_{\mu R}$ ($\bar\nu_{eR}\leftrightarrow\bar\nu'_{\mu L}$), for $Y_e>1/3$ ($Y_e<1/3$), is:
\begin{equation}
 \text{B-res(L)}_e:\qquad
 \left|C\rho\frac{3Y_e-1}{2}\right|
 \simeq \omega_{\rm L}\cos2\theta_{12}\, ,
 \label{eq:dirac-bres-l-active-e}
\end{equation}
independent of the mass ordering. Further conversions of the form
$\nu'_{\mu L}\leftrightarrow\nu_{eR}$ occur when
\begin{equation}
 \text{B-res(L)}_\mu:\qquad
 C\rho\frac{1-Y_e}{2}\simeq \omega_{\rm L}\cos2\theta_{12}\, .
 \label{eq:dirac-bres-l-active-mu}
\end{equation}
Note that conversions of the form $\bar\nu'_{\mu R}\leftrightarrow\bar\nu_{eL}$ do not take place since $0<Y_e<1$.

Concerning  the magnetically-driven resonant  conversions connecting $e$ and $\tau'$ in NO, for $Y_e>1/3$ ($Y_e<1/3$), the condition for resonant conversions of $\nu_{eL}\leftrightarrow\nu'_{\tau R}$ ($\bar\nu_{eR}\leftrightarrow\bar\nu'_{\tau L}$) is
\begin{equation}
 \text{B-res(H)}_e:\qquad
 \left|C\rho\frac{3Y_e-1}{2}\right|
 \simeq |\omega_{\rm H}|\cos2\theta_{13}\, .
 \label{eq:dirac-bres-h-active-e}
\end{equation}
In IO, conversions of the form $\nu_{eL}\leftrightarrow\nu'_{\tau R}$ ($\bar\nu_{eR}\leftrightarrow\bar\nu'_{\tau L}$) require $Y_e<1/3$ ($Y_e>1/3$) instead. Further conversions of the form $\nu'_{\tau L}\leftrightarrow\nu_{eR}$ ($\bar\nu'_{\tau R}\leftrightarrow\bar\nu_{eL}$) take place when
\begin{equation}
 \text{B-res(H)}_\tau:\qquad
 C\rho\frac{1-Y_e}{2}\simeq |\omega_{\rm H}|\cos2\theta_{13}\, 
 \label{eq:dirac-bres-h-active-tau}
\end{equation}
only for NO (IO).

Magnetically-driven resonant conversions  connecting the two non-electron flavors require
\begin{equation}
 \text{B-res(H)}_x:\qquad
 C\rho\frac{1-Y_e}{2}\simeq
 \left|\cos^2\theta_{13}\omega_{\rm H}
 -\left(\cos^2\theta_{12}-\sin^2\theta_{13}\sin^2\theta_{12}\right)\omega_{\rm L}\right|\, .
 \label{eq:dirac-bres-h-mutau-neutrinos}
\end{equation}
In NO, they are $\nu'_{\tau L}\leftrightarrow\nu'_{\mu R}$ and
$\bar\nu'_{\mu R }\leftrightarrow\bar\nu'_{\tau L}$; in IO, they are
$\nu'_{\mu L}\leftrightarrow\nu'_{\tau R}$ and
$\bar\nu'_{\tau R}\leftrightarrow\bar\nu'_{\mu L}$. Both are valid for  $0<Y_e<1$. 

Dirac particles, contrary to Majorana particles, can have intrinsic magnetic moments, which mediate chirality-flipping flavor-conserving conversions. 
In particular, electron-flavor neutrinos and antineutrinos experience magnetically-driven conversions when
\begin{equation}
 \text{B-res}_{ee}:\qquad
 C\rho\frac{3Y_e-1}{2}=0
 \qquad
 \nu_{eL}\leftrightarrow\nu_{eR}\quad
 \bigl(\bar\nu_{eR}\leftrightarrow\bar\nu_{eL}\bigr)\, ,
 \label{eq:dirac-diagonal-ee}
\end{equation}
 for $Y_e=1/3$~\cite{Sasaki:2023sza}.
Resonant conversion of this type for non-electron neutrinos and antineutrinos would require $Y_e =1 $ or $\rho =0$. 

The  MSW and B-res conditions above neglect $V_{\mu\tau}$. This radiative term can induce an additional active $\mu$--$\tau$ resonance~\cite{Esteban-Pretel:2007ncu}, which is negligible for our purposes. 

For Majorana neutrinos, Ref.~\cite{Akhmedov:2003fu} identified an additional
three-flavor resonance (RSFP-E), denoted $\text{B-res}^{*}$ in Ref.~\cite{Manno:2026ikk}. Its derivation relies on the near
degeneracy of three levels in the vicinity of the MSW(H) resonance. For Dirac
neutrinos, we do not find an analogous counterpart. 
Table~\ref{tab:dirac-resonances} summarizes the  magnetically-driven resonant conversions experienced by  Dirac neutrinos.

\begin{table}[t]
\centering
\caption{Matter- and magnetically-driven resonant conversions of Dirac neutrinos for NO and IO.} 
\label{tab:dirac-resonances}
\setlength{\tabcolsep}{12pt}
\renewcommand{\arraystretch}{1.8}
\begin{tabular}{ccc}
\bf Resonance & \bf Normal ordering & \bf Inverted ordering \\
\hline\hline
MSW(L)
& \multicolumn{2}{c}{$\nu_{eL}\leftrightarrow\nu'_{\mu L}$} \\
MSW(H)
& $\nu_{eL}\leftrightarrow\nu'_{\tau L}$
& $\bar\nu_{eR}\leftrightarrow\bar\nu'_{\tau R}$ \\\hline
$\text{B-res(L)}_e$
& \multicolumn{2}{c}{$\begin{matrix}
 \nu_{eL}\leftrightarrow\nu'_{\mu R}\quad(Y_e>1/3)\\
 \bar\nu_{eR}\leftrightarrow\bar\nu'_{\mu L}\quad(Y_e<1/3)
 \end{matrix}$} \\
$\text{B-res(L)}_\mu$
& \multicolumn{2}{c}{$\nu'_{\mu L}\leftrightarrow\nu_{eR}$} \\
$\text{B-res(H)}_e$
& $\begin{matrix}
 \nu_{eL}\leftrightarrow\nu'_{\tau R}\quad(Y_e>1/3)\\
 \bar\nu_{eR}\leftrightarrow\bar\nu'_{\tau L}\quad(Y_e<1/3)
 \end{matrix}$
& $\begin{matrix}
 \nu_{eL}\leftrightarrow\nu'_{\tau R}\quad(Y_e<1/3)\\
 \bar\nu_{eR}\leftrightarrow\bar\nu'_{\tau L}\quad(Y_e>1/3)
 \end{matrix}$ \\
$\text{B-res(H)}_\tau$
& $\nu'_{\tau L}\leftrightarrow\nu_{eR}$
& $\bar\nu'_{\tau R}\leftrightarrow\bar\nu_{eL}$ \\
$\text{B-res(H)}_x$
& $\begin{matrix}
 \nu'_{\tau L}\leftrightarrow\nu'_{\mu R}\\
 \bar\nu'_{\mu R}\leftrightarrow\bar\nu'_{\tau L}
 \end{matrix}$
& $\begin{matrix}
 \nu'_{\mu L}\leftrightarrow\nu'_{\tau R}\\
 \bar\nu'_{\tau R}\leftrightarrow\bar\nu'_{\mu L}
 \end{matrix}$ \\\hline
$\text{B-res}_{ee}$
& \multicolumn{2}{c}{$\nu_{eL}\leftrightarrow\nu_{eR}$,\; $\bar\nu_{eR}\leftrightarrow\bar\nu_{eL}$} \\
\hline\hline
\end{tabular}%
\end{table}

\subsection{A.2~Magnetically-driven resonant conversion in magnetorotational core collapses}

We now assess whether the conditions for resonant conversion are met in magnetorotational core collapses. We extract the  baryon density ($\rho$), electron fraction ($Y_e$), and magnetic field strength ($B$) from our  magnetorotational model with mass of $13 M_\odot$~\cite{Manno:2026ikk,obergaulinger-aloy,Obergaulinger:2021omt} along the polar direction. These quantities depend on the post-bounce time, radius, and emission angle; we omit such dependence  to simplify the notation.

All magnetic moments $(\boldsymbol{\mu}_{\rm D})_{a,b}$ in Eq.~\eqref{eq:dirac-hamiltonians} could be different, depending on the underlying fundamental theory. However, for simplicity,  we consider all to be equal  $(\boldsymbol{\mu}_{\rm D})_{ab}=\mu$, with $a,b=e,\mu',\tau'$. Changes in this assumption would modify the adiabaticity of the different resonances. 

\begin{figure}[t]
\centering
\includegraphics[width=\textwidth]{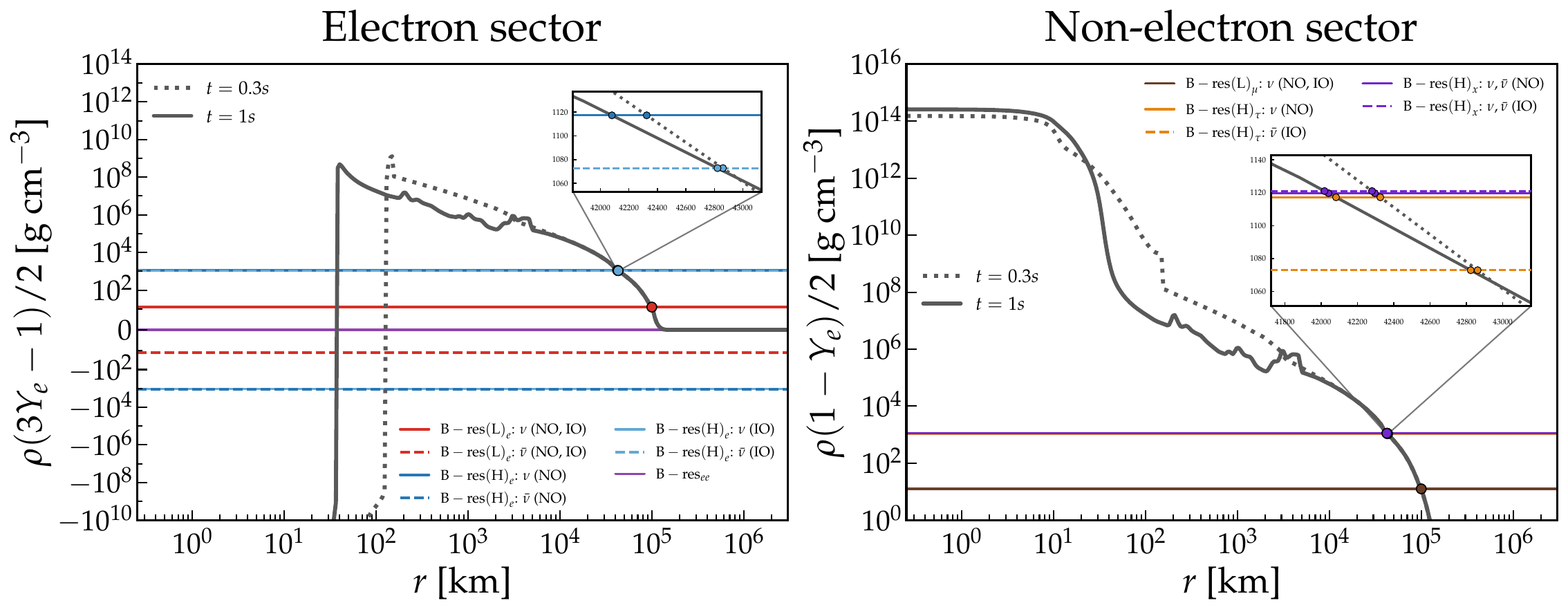}
\caption{Regions of magnetically-induced resonant conversion for Dirac neutrinos, obtained adopting the baryon density and electron fraction profiles of our $13M_\odot$ magnetorotational progenitor extracted along the polar direction at $0.3$~s (dotted gray line) and $1$~s (solid gray line) after bounce. The resonance conditions are evaluated at $E_\nu=14.4\,\mathrm{MeV}$, corresponding to the time-averaged mean energy across all neutrino species along the polar direction~\cite{Manno:2026ikk}. The left and right panels show the resonances in the electron and non-electron sectors, respectively. Colored horizontal lines show the resonance conditions derived in the text. Filled circles mark the outer resonances that are adiabatic, whereas the insets resolve the closely spaced H crossings. The equatorial profiles show a very similar trend at the large radii relevant for these crossings.}
\label{fig:dirac-resonance-locations}
\end{figure}

Figure~\ref{fig:dirac-resonance-locations} shows the profiles $\rho(3Y_e-1)$ and $\rho(1-Y_e)$ along the polar direction at $0.3$~s and $1$~s after bounce. The left panel displays the conditions for the electron sector in Eqs.~\eqref{eq:dirac-bres-l-active-e}, \eqref{eq:dirac-bres-h-active-e}, and \eqref{eq:dirac-diagonal-ee}, which depend on $\rho(3Y_e-1)/2$. The right panel displays the conditions for non-electron neutrino and antineutrino species in Eqs.~\eqref{eq:dirac-bres-l-active-mu}, \eqref{eq:dirac-bres-h-active-tau}, and \eqref{eq:dirac-bres-h-mutau-neutrinos}, which depend on $\rho(1-Y_e)/2$. The horizontal lines represent the  resonance conditions. Where the horizontal line intersects the radial profile, a resonance occurs.  
Because several crossings occur at nearby radii, we highlight the presence of multiple resonances in the figure insets. 

\begin{figure}[t]
    \centering
    \includegraphics[width=\textwidth]{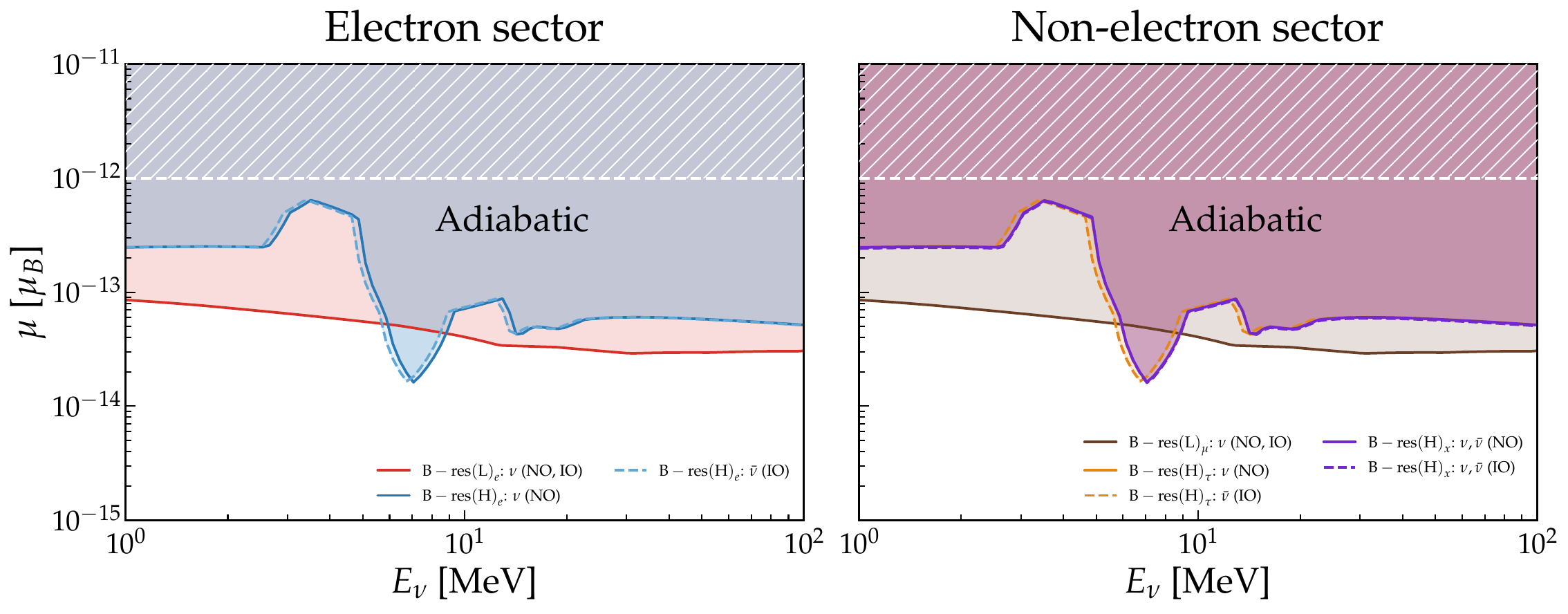}
    \caption{Magnetic moment for which the adiabaticity parameter $\gamma_{ab}=1$ as a function of the neutrino energy, for hydrodynamical quantities and neutrino emission properties extracted along the polar direction  at $1$~s after bounce. The colored lines indicate the  resonances (cf.~Table~\ref{tab:dirac-resonances}) for the crossing occurring further out in Fig.~\ref{fig:dirac-resonance-locations}. For clarity, the electron and non-electron sectors are separated in the left and right panels, respectively (cf.~Fig.~\ref{fig:dirac-resonance-locations}).
    The shaded regions indicate adiabatic conversions, and the hatched
    region indicates magnetic moments above $10^{-12}\,\mu_B$, excluded by the
    most stringent current bounds~\cite{Giunti:2014ixa,ParticleDataGroup:2024cfk}.}
    \label{fig:dirac-adiabaticity-thresholds}
\end{figure}

In the equatorial plane of the magnetorotational model, we find that the type of resonances and the affected radial ranges are comparable to those in the polar direction.
Compared with the Majorana resonances presented in Ref.~\cite{Manno:2026ikk} for the same magnetorotational model, the Dirac resonances occur at larger radii, as summarized in Fig.~\ref{fig:resonance-regions}. 
For each B-res(L) or B-res(H) crossing between an active state and an opposite-chirality state, we denote the corresponding adiabaticity parameter by $\gamma_{ab}$, where $a$ and $b$ identify the two flavors involved, and approximate it as~\cite{Akhmedov:2003fu}
\begin{equation}
 \gamma_{ab}\simeq
 \frac{8E_\nu(\mu B_\perp)^2}{|\Delta m^2_{ij}|}
 \begin{cases}
 \displaystyle
 \left|
 \frac{1}{\rho(3Y_e-1)}
 \frac{d\{\rho(3Y_e-1)\}}{dr}
 \right|^{-1}_{r_{\rm res}}
 & \text{electron sector}\, ,\\[3mm]
 \displaystyle
 \left|
 \frac{1}{\rho(1-Y_e)}
 \frac{d\{\rho(1-Y_e)\}}{dr}
 \right|^{-1}_{r_{\rm res}}
 & \text{non-electron sector}\, .
 \end{cases}
 \label{eq:dirac-adiabaticity}
\end{equation}
Here $\Delta m^2_{ij}=\Delta m^2_{21}$ for B-res(L) and $|\Delta m^2_{ij}|=|\Delta m^2_{31}|$ for B-res(H), and the local quantities are evaluated at the corresponding resonance radius $r_{\rm res}$. The curves in Fig.~\ref{fig:dirac-adiabaticity-thresholds} are obtained by imposing $\gamma_{ab}=1$. Magnetic moments above (below) a given curve correspond to $\gamma_{ab}>1$ ($<1$), describing adiabatic and non adiabatic conversion, respectively.
As shown in Ref.~\cite{Manno:2026ikk} for the same magnetorotational model,  the MSW resonances are always adiabatic. We therefore only need to assess the adiabaticity of magnetically-driven resonant conversions.

The resonant conversions requiring $Y_e<1/3$ occur around $100$~km, where the left panel of Fig.~\ref{fig:dirac-resonance-locations} shows a sharp variation of $\rho(3Y_e-1)/2$. This behaviour follows from the rapid change of $Y_e$.
The diagonal $\text{B-res}_{ee}$ lies in the same region. We therefore treat these inner crossings as non-adiabatic and only retain the outer crossings marked in Fig.~\ref{fig:dirac-resonance-locations}.

For each B-res crossing, we define the Landau-Zener jump probability~\cite{Landau:1932,Zener:1932} as $p_{ab}= \rm exp[-{\pi}/{2}\gamma_{\rm ab}]$. Thus $p_{ab}=0$ ($p_{ab}=1$) corresponds to the adiabatic (non adiabatic) limit. Setting all $p_{ab}=1$ recovers the standard adiabatic MSW only result presented in Eq.~\eqref{eq:msw-only-antinu-fluxes}.

Since the hydrodynamical profiles evolve with time, so do the resonance locations. Differences also arise for neutrinos being emitted along different directions  from the 3D magnetorotational model, but these variations do not appreciably affect our conclusions. For the level crossing schemes and flux calculation, we adopt the polar profile at $1$~s after bounce as a representative case.

For our magnetorotational model, at the representative energy $E_\nu=14.4$~MeV, the threshold magnetic moments are  $\simeq 10^{-14}$--$10^{-13}\,\mu_B$, and all the outer B-res crossings are adiabatic for $\mu\gtrsim5\times10^{-14}\,\mu_B$. Above the relevant thresholds, the magnetically-driven resonant conversions of Dirac neutrinos are adiabatic, similar to the Majorana case~\cite{Manno:2026ikk}. These values lie below the most stringent astrophysical bounds on neutrino magnetic moments, $\mu\lesssim\mathcal O(10^{-12})\,\mu_B$~\cite{Giunti:2014ixa,ParticleDataGroup:2024cfk}.

\begin{figure*}[t]
\centering
\includegraphics[width=\textwidth]{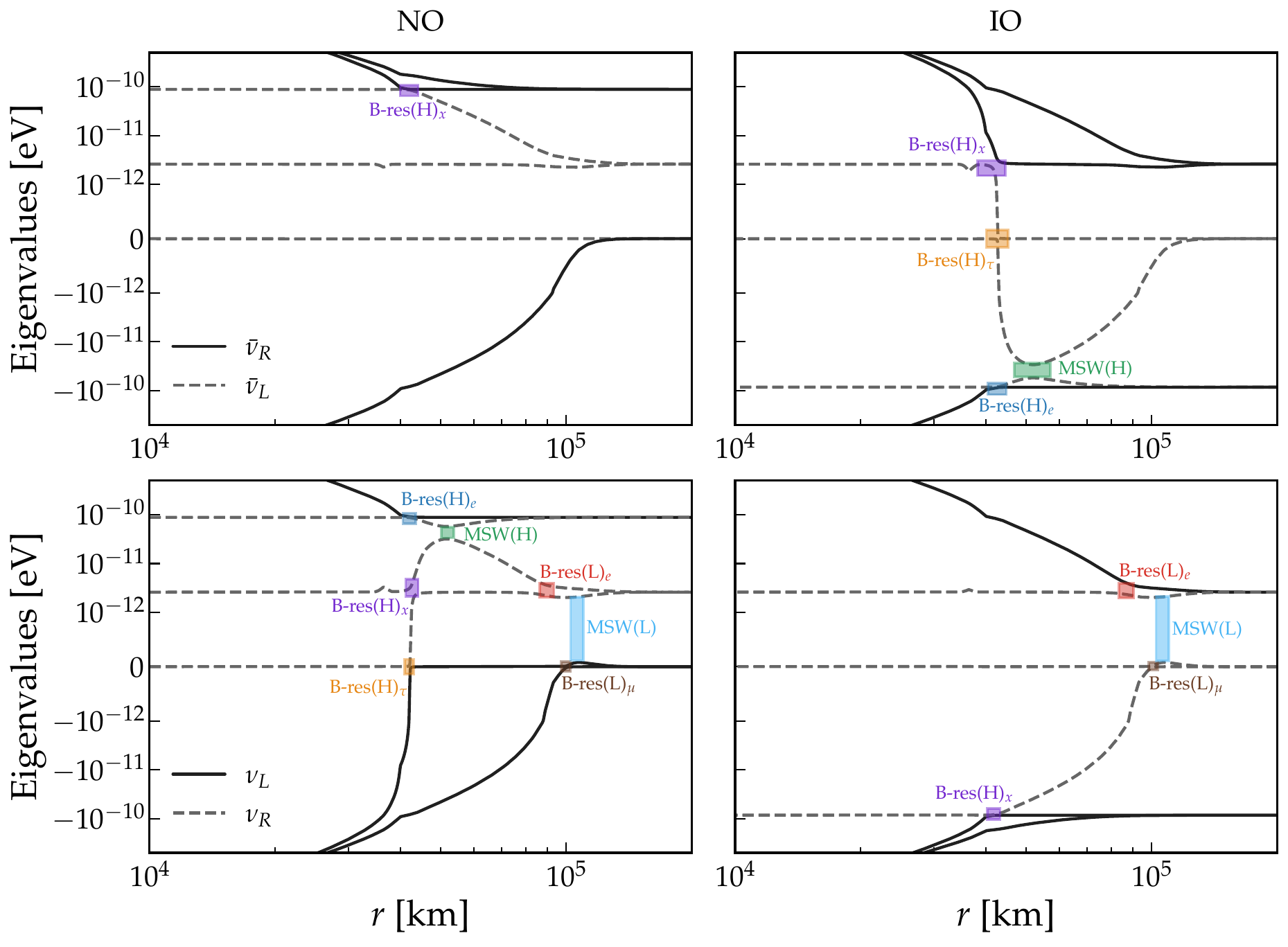}
\caption{Level-crossing schemes for Dirac antineutrinos (top) and neutrinos (bottom) for the polar direction of the $13M_\odot$ magnetorotational model,  at $t=1\,\mathrm{s}$ after bounce, and $E_\nu=14.4\,\mathrm{MeV}$ and $\mu=10^{-13}\mu_B$. The selected energy represents the time-averaged mean energy across all neutrino species along the polar direction~\cite{Manno:2026ikk}. The level-crossing structure is similar at $t=0.3\,\mathrm{s}$ and along the equatorial direction. The left and right columns show NO and IO, respectively. Solid and dashed branches represent  active and opposite-chirality states. The  boxes in color highlight the crossings that are potentially adiabatic in the magnetorotational profiles and show the radial regions in which they occur. Compared with the Majorana case~\cite{Manno:2026ikk}, these crossings occur predominantly at larger radii.}
\label{fig:dirac-level-schemes}
\end{figure*}

Figure~\ref{fig:dirac-level-schemes} presents the level-crossing schemes associated with the resonances experienced by Dirac particles, namely the radial evolution of the Hamiltonian eigenvalues, with antineutrinos in the upper row and neutrinos in the lower row.
The eigenvalues are shown in physical units to identify the radial positions of the crossings and the sequence in which they occur. Solid black (dashed gray) branches are initialized in the active (opposite-chirality) sector. Thus, they correspond to $\bar\nu_R$ ($\bar\nu_L$) for antineutrinos and to $\nu_L$ ($\nu_R$) for neutrinos.

\subsection{A.3~Fluxes expected at Earth}
Throughout this section, $\varphi_{\nu_\alpha}^0(E_\nu)$ and
$\varphi_{\nu_\alpha}(E_\nu)$, with $\alpha=e,\mu,\tau$, denote the differential number flux at production and after flavor conversion, respectively. The same notation applies to antineutrinos.
These spectra are evaluated for $0\leq E_\nu\leq 240$~MeV.
We obtain the fluxes at Earth relying on the level crossing schemes in Fig.~\ref{fig:dirac-level-schemes}.
During propagation to Earth, the mass eigenstates lose coherence. The flux observed in flavor $\alpha=e,\mu,\tau$ is therefore the incoherent sum $\varphi_{\nu_\alpha}=\sum_{i=1}^{3}|U_{\alpha i}|^2\varphi_{\nu_i}$, where $\varphi_{\nu_i}$ is the flux carried by the mass eigenstate $\nu_i$; the same relation holds for antineutrinos.

The observable flux of $\bar\nu_e$'s in NO is
\begin{equation}
 \begin{split}
 \varphi_{\bar\nu_{eR}}^{\mathrm{D,NO}}&=
 |U_{e1}|^2\varphi^0_{\bar\nu_{eR}}
 +|U_{e2}|^2\left[
 p_{\mu\tau}\varphi^0_{\bar\nu'_{\mu R}}
 +(1-p_{\mu\tau})\varphi^0_{\bar\nu'_{\tau L}}
 \right] +|U_{e3}|^2\varphi^0_{\bar\nu'_{\tau R}}\, ,
 \end{split}
 \label{eq:dirac-no-antinu-general}
\end{equation}
where $p_{\mu\tau}$ denotes the jump probability at $\bar\nu'_{\mu R}\leftrightarrow\bar\nu'_{\tau L}$. The opposite-chirality component is not emitted at the source
($\varphi^0_{\bar\nu'_{\tau L}}=0$). For magnetic moments above the range quoted above, $p_{\mu\tau}=0$ and Eq.~\eqref{eq:dirac-no-antinu-general} reduces to  Eq.~\eqref{eq:fluxes-dirac}, upon identifying $\varphi_{\bar\nu_{eR}}=\varphi_{\bar\nu_e}$, $\varphi^0_{\bar\nu_{eR}}=\varphi^0_{\bar\nu_e}$, and $\varphi^0_{\bar\nu'_{\tau R}}=\varphi^0_{\nu_x}$.

For IO, $p_{e\tau}$, $p_{\tau e}$, and $p_{\tau\mu}$ denote jump probabilities at $\bar\nu_{eR}\leftrightarrow\bar\nu'_{\tau L}$, $\bar\nu'_{\tau R}\leftrightarrow\bar\nu_{eL}$, and $\bar\nu'_{\tau R}\leftrightarrow\bar\nu'_{\mu L}$, respectively, and the observable flux is 
\begin{equation}
 \begin{split}
 \varphi_{\bar\nu_{eR}}^{\mathrm{D,IO}}&=
 |U_{e1}|^2\Big[
 p_{\tau e}\big(
 p_{\tau\mu}\varphi^0_{\bar\nu'_{\tau R}}
 +(1-p_{\tau\mu})\varphi^0_{\bar\nu'_{\mu L}}
 \big)+(1-p_{\tau e})\varphi^0_{\bar\nu_{eL}}
 \Big] +|U_{e2}|^2\varphi^0_{\bar\nu'_{\mu R}} 
+|U_{e3}|^2\Big[
 p_{e\tau}\varphi^0_{\bar\nu_{eR}}
 +(1-p_{e\tau})\varphi^0_{\bar\nu'_{\tau L}}
 \Big]\, .
 \end{split}
 \label{eq:dirac-io-antinu-general}
\end{equation}
At the source, $\varphi^0_{\bar\nu_{eL}}=\varphi^0_{\bar\nu'_{\mu L}}= \varphi^0_{\bar\nu'_{\tau L}}=0$ and $\varphi^0_{\bar\nu'_{\mu R}}=\varphi^0_{\bar\nu'_{\tau R}}=\varphi^0_{\nu_x}$. Hence, for adiabatic magnetic conversion, $p_{e\tau}=p_{\tau e}=p_{\tau\mu}=0$, the result reduces to the expression in Eq.~\eqref{eq:fluxes-dirac}.

For completeness, we also report the flux of Dirac electron neutrinos.  In NO, $p_{e\mu}$, $p_{e\tau}$, $p_{\mu e}$, $p_{\mu\tau}$, and $p_{\tau e}$ are the jump probabilities for $\nu'_{\mu L}\leftrightarrow\nu_{eR}$, $\nu'_{\tau L}\leftrightarrow\nu_{eR}$, $\nu_{eL}\leftrightarrow\nu'_{\mu R}$, $\nu'_{\tau L}\leftrightarrow\nu'_{\mu R}$, and $\nu_{eL}\leftrightarrow\nu'_{\tau R}$, respectively. The observable flux is
\begin{align}
 \varphi_{\nu_e}^{\mathrm{D,NO}}&=
 |U_{e1}|^2\left\{
 p_{e\mu}\varphi^0_{\nu'_{\mu L}}
 +(1-p_{e\mu})\left[
 (1-p_{e\tau})\varphi^0_{\nu'_{\tau L}}
 +p_{e\tau}\varphi^0_{\nu_{eR}}
 \right]
 \right\}
 +|U_{e2}|^2\Biggl\{
 p_{\mu e}\Bigl[
 p_{\mu\tau}\Bigl[
 p_{e\tau}\varphi^0_{\nu'_{\tau L}}
 +(1-p_{e\tau})\varphi^0_{\nu_{eR}}
 \Bigr]
 \nonumber\\
 &\qquad
 +(1-p_{\mu\tau})\varphi^0_{\nu'_{\mu R}}
 \Bigr]
 +(1-p_{\mu e})\Bigl[
 (1-p_{\mu\tau})\Bigl[
 p_{e\tau}\varphi^0_{\nu'_{\tau L}}
 +(1-p_{e\tau})\varphi^0_{\nu_{eR}}
 \Bigr]
 +p_{\mu\tau}\varphi^0_{\nu'_{\mu R}}
 \Bigr]
 \Biggr\}
 \nonumber\\
 &\qquad+|U_{e3}|^2\left[
 p_{\tau e}\varphi^0_{\nu_{eL}}
 +(1-p_{\tau e})\varphi^0_{\nu'_{\tau R}}
 \right]\, .
 \label{eq:dirac-no-nu-general}
\end{align}
The opposite-chirality states are not populated at production: $\varphi^0_{\nu_{eR}}=\varphi^0_{\nu'_{\mu R}}=\varphi^0_{\nu'_{\tau R}}=0$. We also use $\varphi^0_{\nu'_{\mu L}}=\varphi^0_{\nu'_{\tau L}}=\varphi^0_{\nu_x}$ and $\varphi^0_{\nu_{eL}}=\varphi^0_{\nu_e}$. If all magnetic resonances are adiabatic, $p_{e\mu}=p_{e\tau}=p_{\mu e}=p_{\mu\tau}=p_{\tau e}=0$, and Eq.~\eqref{eq:dirac-no-nu-general} reduces to $\varphi_{\nu_e}^{\mathrm{D,NO}}=|U_{e1}|^2\varphi^0_{\nu_x}$.

In IO, $p_{\tau\mu}$, $p_{\mu e}$, and $p_{e\mu}$ represent the jump probabilities for $\nu'_{\mu L}\leftrightarrow\nu'_{\tau R}$, $\nu_{eL}\leftrightarrow\nu'_{\mu R}$, and $\nu'_{\mu L}\leftrightarrow\nu_{eR}$. Therefore, we have
\begin{align}
 \varphi_{\nu_e}^{\mathrm{D,IO}}&=
 |U_{e1}|^2\left\{
 p_{e\mu}\left[
 p_{\tau\mu}\varphi^0_{\nu'_{\mu L}}
 +(1-p_{\tau\mu})\varphi^0_{\nu'_{\tau R}}
 \right]
 +(1-p_{e\mu})\varphi^0_{\nu_{eR}}
 \right\}+|U_{e2}|^2\left[
 p_{\mu e}\varphi^0_{\nu_{eL}}
 +(1-p_{\mu e})\varphi^0_{\nu'_{\mu R}}
 \right]+|U_{e3}|^2\varphi^0_{\nu'_{\tau L}}.
 \label{eq:dirac-io-nu-general}
\end{align}
At production, the opposite-chirality neutrino fluxes vanish, while $\varphi^0_{\nu'_{\mu L}}=\varphi^0_{\nu'_{\tau L}}=\varphi^0_{\nu_x}$ and $\varphi^0_{\nu_{eL}}=\varphi^0_{\nu_e}$. If all magnetic resonances are adiabatic, $p_{\tau\mu}=p_{\mu e}=p_{e\mu}=0$, and Eq.~\eqref{eq:dirac-io-nu-general} becomes $\varphi_{\nu_e}^{\mathrm{D,IO}}=|U_{e3}|^2\varphi^0_{\nu_x}$.
The DSNB could be observed in $\nu_e$'s in DUNE~\cite{DUNE:2025sjq}. However, we have verified that including DUNE in the combined analysis does not appreciably improve the Dirac-Majorana sensitivity relative to the Hyper-Kamiokande-Gd+JUNO result shown in Fig.~\ref{fig:sensitivity-majorana-dirac}.

\section{B.~MSW conversion of Dirac and Majorana neutrinos}
The fluxes discussed above correspond to the regime in which the magnetically-induced resonant  conversions are adiabatic. As shown in Fig.~\ref{fig:dirac-adiabaticity-thresholds}, this occurs when the magnetic moment is above the threshold $10^{-14}\,\mu_B$. For smaller magnetic moments,  flavor conversion is non-adiabatic; however, the standard MSW H and L resonances are  adiabatic. In this case, the  $\bar\nu_e$ fluxes are then common
to Dirac and Majorana neutrinos and read as
\begin{equation}
 \begin{aligned}
 \varphi_{\bar\nu_e}^{\mathrm{MSW,NO}}
 &=
 |U_{e1}|^2\varphi_{\bar\nu_e}^0
 +( |U_{e2}|^2+|U_{e3}|^2 )\varphi_{\nu_x}^0\, ,\\
 \varphi_{\bar\nu_e}^{\mathrm{MSW,IO}}
 &=
 |U_{e3}|^2\varphi_{\bar\nu_e}^0
 +( |U_{e1}|^2+|U_{e2}|^2 )\varphi_{\nu_x}^0\, .
 \end{aligned}
 \label{eq:msw-only-antinu-fluxes}
\end{equation}
Therefore, if only MSW conversion takes place, it is not possible to distinguish between Dirac and Majorana particles.

\begin{figure}[t]
    \centering
    \includegraphics[width=0.48\textwidth]{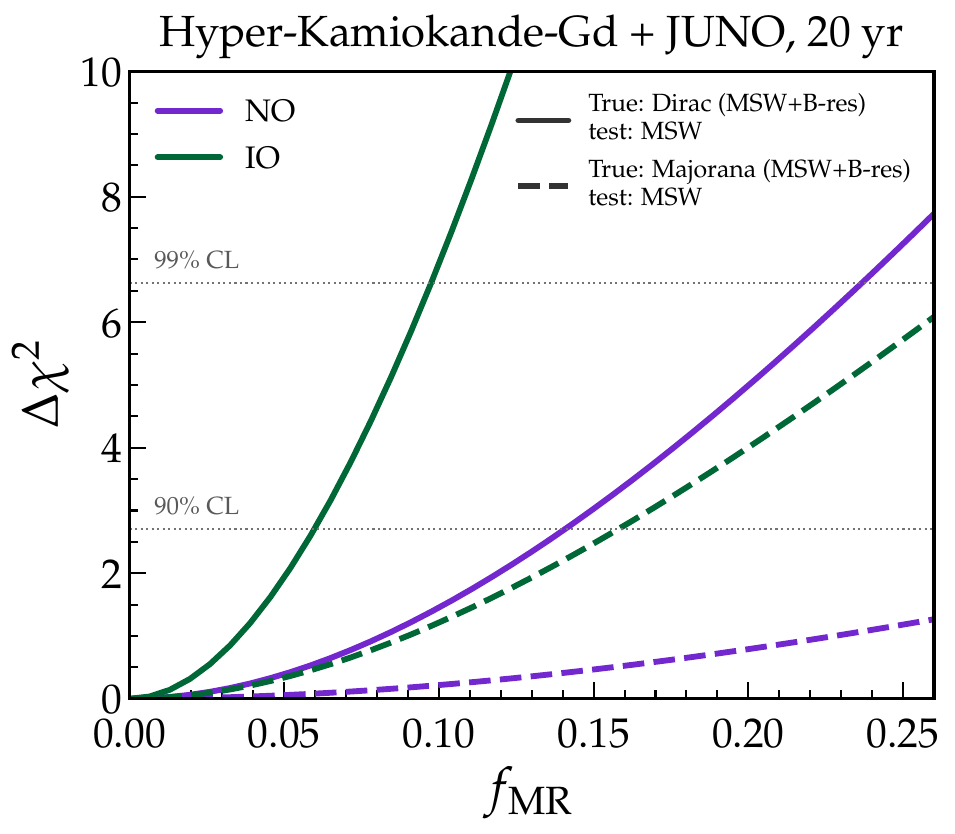}
    \caption{Expected sensitivity to distinguish between the case with large neutrino magnetic moments (where magnetically-induced resonant  conversions and MSW flavor conversion occurs) from the scenario with small magnetic moment (where only MSW conversion takes place), using $20$~yr of combined Hyper-Kamiokande-Gd and JUNO data. Purple and green lines denote NO and IO, respectively. Solid (dashed) lines assume the Dirac (Majorana) MSW+B-res prediction as the true hypothesis and the MSW only prediction as the test hypothesis. The latter is independent of the Dirac or Majorana nature of neutrinos. The horizontal dotted lines mark the $90\%$ and $99\%$ confidence levels.     }
    \label{fig:sensitivity-magnetic-msw}
\end{figure}

Figure~\ref{fig:sensitivity-magnetic-msw} shows the comparison between the regimes of large and small magnetic moments, i.e.~magnetically-induced resonant  conversions $+$ MSW vs.~MSW-only scenarios, for both mass orderings. The case of  Dirac particles with large magnetic moments can be distinguished from both the one with small magnetic moments (i.e., MSW only) and the one for Majorana neutrinos with large magnetic moments at $90\%$ and $99\%$ confidence, in the allowed range of $f_{\rm MR}$. If neutrinos were Majorana, in NO it would not be possible to gather significant evidence to discriminate standard MSW conversions from the scenario with magnetically-induced resonant  conversions and MSW. However,   a hint of confidence could be reached in $90\%$ for $f_{\rm MR}\gtrsim0.16$ in IO. 

\end{document}